\documentclass[twocolumn,           % Format : preprint, twocolumn
               showpacs,            % Pacs : showpacs, noshowpacs
               preprintnumbers,     % Preprint: preprintnumbers,
               aps,                 % Society: ...
               prd,                 % Journal Style : pra, prb, prc, prd, pre,
               a4paper,             % Size : a4paper, ...
               superscriptaddress,  % Affiliation (Title) : groupedaddress,
               nofootinbib,         % Footnote: footinbib, nofootinbib
               tightenlines,        % Remove additional spaces in a line
               floats,floatfix      % Floating pictures and tables
               ]{revtex4}

\usepackage{graphicx}
\usepackage{latexsym}
\usepackage{amsmath,amssymb}        % amssymb includes amsfonts
\usepackage[draft=false]{hyperref}

\input colordvi

\begin{document}

%%--- DRAFTCOPY --------------------------------
%% Prints a large "DRAFT" diagonally across each page
%% Does not show up in TeXview
%% \typeout{Prints "DRAFT" on each page; does not show in TeXView}
%% \special{!userdict begin /bop-hook{gsave 200 30 translate
%% 65 rotate /Times-Roman findfont 216 scalefont setfont
%% 0 0 moveto 0.90 setgray (DRAFT) show grestore}def end}
%%------------------------------------------------

%======================================%
%<<<<<<<<<<<< TITLE PAGE >>>>>>>>>>>>>>%
%======================================%

\title{Structure formation constraints on the Jordan--Brans--Dicke theory}
\author{Viviana Acquaviva}
\affiliation{SISSA/ISAS, Via Beirut 4, 34014 Trieste, Italy}
\affiliation{INFN, Sezione di Trieste, Via Valerio 2, 34127 Trieste, Italy}
\affiliation{Astronomy Centre, University of Sussex, Brighton BN1 9QH, United 
Kingdom}
\author{Carlo Baccigalupi}
\affiliation{SISSA/ISAS, Via Beirut 4, 34014 Trieste, Italy}
\affiliation{INFN, Sezione di Trieste, Via Valerio 2, 34127 Trieste, Italy}
\author{Samuel M.~Leach}
\affiliation{SISSA/ISAS, Via Beirut 4, 34014 Trieste, Italy}
\affiliation{D\'epartement de Physique Th\'eorique, Universit\'e de Gen\`eve,
24 quai Ernest Ansermet, CH-1211 Gen\`eve 4, Switzerland}
\author{Andrew R.~Liddle}
\affiliation{Astronomy Centre, University of Sussex, Brighton BN1 9QH, United 
Kingdom}
\author{Francesca Perrotta}
\affiliation{SISSA/ISAS, Via Beirut 4, 34014 Trieste, Italy}
\affiliation{INFN, Sezione di Trieste, Via Valerio 2, 34127 Trieste, Italy}
\date{\today}
\pacs{04.50.+h, 98.80.Es \hfill astro-ph/0412052}
\preprint{astro-ph/0412052}

%======================================%
%<<<<<<<<<<<<< ABSTRACT >>>>>>>>>>>>>>>%
%======================================%

\begin{abstract}
We use cosmic microwave background data from WMAP, ACBAR, VSA and CBI, and 
galaxy power spectrum data from 2dF, to constrain flat cosmologies based on the
Jordan--Brans--Dicke theory, using a Markov Chain Monte 
Carlo approach.
Using a parametrization based on $\xi=1/4\omega$, and performing an
exploration in the range $\ln \xi \in [-9,3]$, we obtain a $95\%$
marginalized probability bound of $\ln\xi<-6.2$, corresponding to
a $95\%$ marginalized probability lower bound on the Brans--Dicke
parameter $\omega>120$.
\end{abstract}

\maketitle

%======================================%
%<<<<<<<<<<<<<< ARTICLE >>>>>>>>>>>>>>>%
%======================================%

\section{Introduction}

Jordan--Brans--Dicke (JBD) theory \cite{JBD,Will} is the simplest extended 
theory of gravity, depending on one additional parameter, the Brans--Dicke 
coupling $\omega$, as compared to General Relativity. 
As Einstein's theory is recovered in the limit $\omega \rightarrow \infty$, 
there will always be viable JBD theories as long as General Relativity remains 
so too. As such, it acts as a laboratory for quantifying how accurately the 
predictions of General Relativity stand up against observational tests. The most 
stringent limits are derived from radar timing experiments within our Solar 
System, with measurements using the Cassini probe \cite{Cassini} now giving 
a two-sigma lower limit $\omega > 40,000$ (improving pre-existing limits 
\cite{Willlive} by an order of magnitude).

With precision cosmological data now available, particularly on cosmic microwave 
background (CMB) anisotropies from the Wilkinson Microwave Anisotropy Probe 
(WMAP) \cite{wmap}, it has become feasible to obtain complementary constraints 
from the effect of modified gravity on the structure formation process, as 
suggested in Ref.~\cite{LMB}. That paper focussed on the way 
that $\omega$ alters the Hubble scale at matter--radiation equality, which is a 
scale imprinted on the matter power spectrum, in an attempt to identify how 
large an effect can be expected. Subsequently, the expected total intensity and 
polarization 
microwave anisotropy spectra in the JBD theory were computed,
and a forecast of the sensitivity to $\omega$ of data 
from the WMAP and Planck satellites carried out exploiting a
Fisher matrix approach \cite{ck}.

In this paper we make a comprehensive comparison of predictions of the JBD 
theory to
current observational data, using WMAP and other CMB data plus the galaxy 
power spectrum as measured by the two-degree field (2dF) galaxy redshift survey. 
We define JBD models 
in terms of eight parameters, which are allowed to vary simultaneously.  Our 
paper is closest in 
spirit to work by Nagata et al.~\cite{NCS}, who considered a more general model, 
the harmonic attractor model, which includes JBD as a special case. However  
their dataset compilation was restricted to the WMAP temperature power spectrum.

The constraint we will obtain is not competitive with the very stringent solar 
system bound given above (though the analysis of Ref.~\cite{ck}
indicates that a limit as high as 3000 might eventually be reached by the
measurements of the Planck satellite), but it is complementary in that it 
applies 
on a 
completely different length and time scale. Such constraints can therefore 
still be of interest in general scalar--tensor theories where $\omega$ is 
allowed to vary; for instance Nagata et al.~\cite{NCS} find that in some 
parameter regimes of the harmonic attractor model the cosmological constraint is 
stronger than the Solar System one. In that regard, our result is most 
comparable to cosmological 
constraints imposed on $\omega$ from nucleosynthesis, which give only a weak 
lower limit of $\omega > 32$ \cite{nucleo}. 

\section{Formalism}

\subsection{Background cosmology}

The Lagrangian for the JBD theory is
\begin{equation}
{\cal L} = \frac{m_{{\rm Pl}}^2}{16\pi} \left( \Phi R - \frac{\omega}{\Phi} 
\partial_\mu \Phi \partial^\mu \Phi \right) + {\cal L}_{{\rm matter}} \,,
\end{equation}
where the Brans--Dicke coupling $\omega$ is a constant, and $\Phi (t)$ is the 
Brans--Dicke (BD) field whose
present value must give the observed gravitational coupling. We have included 
factors of 
$m_{{\rm Pl}}$ to define $\Phi$ as dimensionless.

The equations for a spatially-flat Friedmann universe are 
\cite{JBD,Will,Wein} 
\begin{eqnarray}
\left( \frac{\dot a}a\right) ^2+\frac{\dot a}a\,\frac{\dot \Phi }\Phi  
        &=& \frac{\omega}{6} \left( \frac{\dot \Phi }\Phi \right)^2
        +\frac{8\pi}{3m_{{\rm Pl}}^2 \, \Phi} \rho \,;  \label{bdfield} \\
\ddot \Phi +3\frac{\dot a}a\,\dot \Phi  &=&\frac{8\pi }{(2\omega +3)m_{{\rm
        Pl}}^2} \left(\rho -3p\right) \,,  \label{bdfield2}
\end{eqnarray}
where $a(t)$ is the
cosmological scale factor, and $\rho$
and $p$ are the energy density and pressure summed over all types of material in 
the Universe. 

The Universe is assumed to contain the same ingredients as the WMAP concordance 
model \cite{wmap}, namely dark energy, dark matter, baryons, photons and 
neutrinos. We make the simplifying assumptions of spatial flatness, dark energy 
in the form of a pure cosmological constant, and effectively massless neutrinos 
whose density is related to that of photons by the usual thermal argument. The 
present value of $\Phi$ must correctly reproduce the strength of gravity seen in 
Cavendish-like experiments, which requires \cite{Will}
\begin{equation}
\Phi_0 = \frac{2\omega +4}{2\omega +3} \,,
\end{equation}
where here and throughout a subscript `0' indicates present value. We will 
assume that the value of $\Phi_0$ in our Solar System is representative of the 
Universe as a whole, though this may not be absolutely accurate \cite{CMB}. We 
also assume 
that the initial perturbations are given by a power-law adiabatic perturbation 
spectrum.

\begin{figure}[t]

\includegraphics[width=\linewidth]{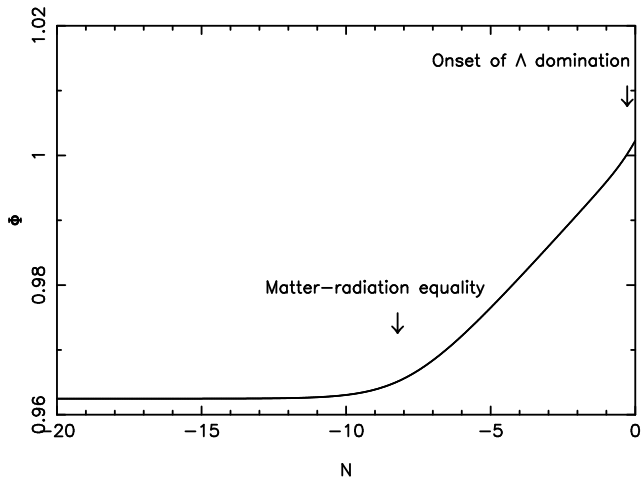}\\
\caption[bdevolve]{\label{bdevolve} Evolution of the BD field from 
early in radiation domination to the present. It is just possible to see the 
evolution of $\Phi$ increase as $\Lambda$ domination sets in. The cosmological 
parameters are $\omega = 200$, $H_0 = 72$, and $\rho_{{\rm m},0} = 0.3$ in units 
of the standard cosmology critical density.}

\end{figure}

When the Universe is dominated by a single fluid there are a variety of analytic 
solutions known \cite{Nar}, where $\Phi$ is typically constant during a 
radiation era, slowly increasing during a matter era, and then more swiftly 
evolving as dark energy domination sets in. However we need solutions spanning 
all three eras and so will solve the equations numerically, for which we use the 
integration variable $N \equiv \ln a/a_0$. An example of the evolution is shown 
in Fig.~\ref{bdevolve}.

The basic parameter set we use to build our cosmological models contains the 
following parameters
\begin{tabbing}
\hspace*{0.5cm} \= $\omega$~~~~~ \= Brans--Dicke coupling\\
\> $H_0$ \> present Hubble parameter $[{\rm km}\, {\rm s}^{-1}{\rm Mpc}^{-1}]$\\
\> $\rho_{{\rm B}}$ \> baryon density\\
\> $\rho_{{\rm C}}$ \> cold dark matter density\\
\> $A_{{\rm S}}$ \> curvature perturbation amplitude\\
\> $n_{{\rm S}}$ \> perturbation spectral index\\
%\> ${\mathcal Z}$ \> reionization optical depth, ${\mathcal Z} \equiv 
\> $\tau$ \> reionization optical depth \\
\> $b$ \> galaxy bias parameter, $P_{{\rm gg}}/P_{{\rm mm}}$
\end{tabbing}
where $b^2=P_{{\rm gg}}/P_{{\rm mm}}$ is the ratio of the (observed)
galaxy power spectrum to the (calculated) matter power spectrum.
Other parameters are fixed by the assumptions above, and the radiation energy 
density is taken as fixed by the direct observation of the CMB temperature $T_0 
= 2.725$K \cite{Math99}. 

An important subtlety that must be taken into account is that the extra terms in 
Eq.~(\ref{bdfield}), plus the Cavendish-like correction to the present value of 
$\Phi$, means that the usual relation between the Hubble parameter and density, 
used to define the critical density and hence density parameters, no longer 
applies. Generically, the extra terms require an increase in the present value 
of $\rho$ to give the same expansion rate, the correction being of order 
$1/\omega$. Because of this subtlety, we define 
the density parameters $\Omega_{\rm B,C}$ by dividing by the critical
density for the standard  cosmology, meaning that the density
parameters don't quite sum to one for a spatially-flat model. 

Operationally, we proceed as follows. We seek a background evolution 
corresponding to a particular value of $h=H_0/100$ and of the present physical 
matter 
density. We can assume the initial velocity of the BD field $\dot{\Phi}$ is zero 
deep in the radiation era, which leaves us two parameters, the early time value 
of $\Phi$ and the value of the cosmological constant, to adjust in order to 
achieve the required values. This is a uniquely-defined problem, with the 
necessary values readily found via an iterative shooting method. 

\subsection{Perturbation evolution}

We carry out the evolution of density perturbations using a modified version of 
the code {\sc defast}, based on {\sc cmbfast} \cite{SZ} and originally written 
to study quintessence scenarios where the dark energy scalar field is minimally 
\cite{PB} or non-minimally \cite{PBM} coupled to the Ricci scalar. The 
architecture of {\sc defast} is based on the version 4.0 of {\sc cmbfast}, 
although 
there has been a progressive code fork in the subsequent versions. 
{\sc defast} takes as input the parameter set described in the previous 
subsection, and returns the microwave anisotropy spectra (for temperature and 
polarization) and the matter power spectrum. A dynamical and fluctuating scalar 
field, playing the role of the dark energy and/or the BD field, is 
included 
into the analysis together with the other cosmological components, following the 
existing general scheme \cite{hwang}. 

In order to bring the model description into the formalism used by {\sc defast}, 
we redefine the BD field and coupling according to
\begin{equation}
\phi^2 = \omega \,\Phi \frac{m_{{\rm Pl}}^2}{2\pi} \quad ; \quad \xi = 
\frac{1}{4\omega} \,,
\end{equation}
which brings the Lagrangian into the form
\begin{equation}
{\cal L} = \frac{1}{2} \xi \phi^2 R - \frac{1}{2} \partial_\mu \phi \partial^\mu 
\phi + {\cal L}_{{\rm matter}} \,,
\end{equation}
where $\phi$ is now a canonical scalar field non-minimally coupled to gravity. 
We implement the cosmological constant in the code by giving $\phi$ a constant 
potential energy.

Our calculations include the effect of perturbations, with the initial 
perturbations in $\phi$ fixed by the requirement of adiabaticity. The 
correction to the background expansion rate from the dynamics of $\phi$
is the most relevant effect on the CMB power spectrum, appearing as 
a projection plus a correction to the Integrated Sachs--Wolfe (ISW)
effect, as discussed in detail in Ref.~\cite{PBM}.

\subsection{Data analysis}

The data we use are taken from WMAP \cite{wmap_data} and the 2dF
galaxy redshift survey expressed as 32 bandpowers in the range 
$0.02<k<0.15h^{-1}$Mpc \cite{2df}. In order to incorporate the 2dF data,
the galaxy bias parameter $b$ is taken to be a free parameter for which
the analytic marginalization scheme of Ref.~\cite{bridle} can be applied.
We also consider the effect of including the high-$\ell$ CMB data
from VSA \cite{vsa}, CBI \cite{cbi}, ACBAR \cite{acbar}.

Our present analysis does not include supernovae data. Inclusion 
of the modification to the luminosity distance from $\omega$ would be 
straightforward. However the variation of the gravitational coupling $G$ means 
that supernovae can no 
longer be assumed to be standard candles, and Ref.~\cite{Getal} suggests that 
the effect from varying $G$ dominates. Further, inclusion of supernovae data may 
be particularly susceptible to the possibility that the local value of $\Phi$ in 
the vicinity of the supernova may not match the global cosmological value 
\cite{CMB}. Nevertheless, it would be interesting to investigate robust methods 
for including such data, also in connection with alternative
observational strategies \cite{Boiss}.

We carry out the data analysis using the now-standard Markov Chain Monte Carlo
posterior sampling technique, by modifying the June 2004 version of the {\sc
CosmoMC} program \cite{LB} to call {\sc defast} to obtain the spectra.  {\sc
CosmoMC} computes the likelihood of the returned model and assembles a set of
samples from the posterior distribution.  We take full advantage of {\sc
CosmoMC}'s MPI capabilities by running the code across 19 Sun V60x Xeon 2.8GHz
processors.  The Metropolis--Hastings algorithm is run at a temperature of 1.3
in order to better sample the non-Gaussian direction of our posterior
distribution which results from the degeneracy between $H_0$ and $\ln\xi$, both
of which have a strong effect on the angular diameter distance.  The final
chains are then cooled and importance sampled \cite{LB}.  It can be noted that
for the purposes of posterior sampling, we have parametrized the JBD cosmology
using $\ln\xi\equiv -\ln 4\omega$ simply because it is more straightforward to
obtain the samples we need, while simultaneously suppressing the possibility of
jumping to regions with $\omega<1$.  Specifically, we use a flat prior on
$\ln\xi \in [-9,-3]$ where the lower cutoff has been adjusted to the point where
the likelihood function is no longer sensitive to the effect of varying the
Brans--Dicke parameter and the $\Lambda{\rm CDM}$ model is thereby recovered.
As usual, this Jeffreys prior, which is defined here as a flat prior on the
logarithm of a parameter of unknown scale, has the interesting property of
invariance under scale reparametrizations \cite{HJ}.  For this reason it serves
as a reasonable substitute for working with a more desirable physical parameter
which could be identified to isolate and give a linear response in the ISW
effect, mainly responsible for the upper bound on $\ln\xi$.

The optical depth $\tau$ is parametrized using ${\mathcal Z}=\exp{[-2\tau]}$,
where ${\mathcal Z}^{1/2}$ is the fraction of photons that remain unscattered
through reionization, since the combination $A_{{\rm S}}{\mathcal Z}$ is well
constrained by the CMB.

The results that we present are based on around 100,000 raw posterior samples, 
and while the basic constraints can be derived with significantly
fewer samples, this large number assures more robust constraints on
the derived parameter $\omega$  when we use importance sampling in
order to adjust for the change in prior density \cite{LB}.

\begin{figure}[t]
\includegraphics[width=\linewidth]{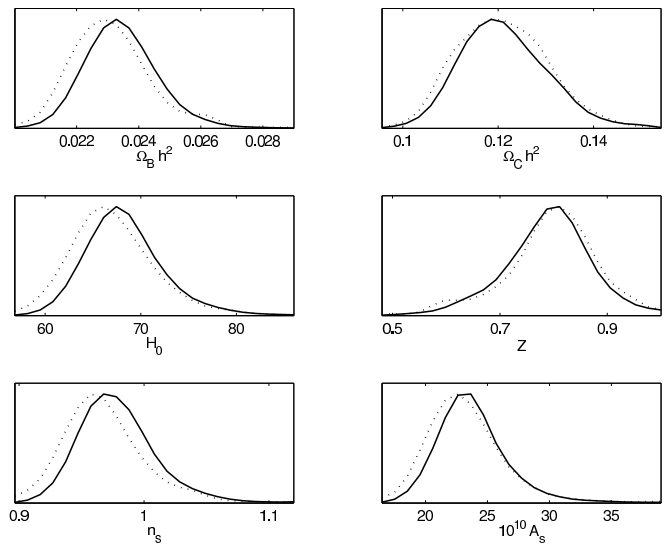}\\
\caption[1D_constraints]{\label{1D_constraints} Marginalized 1D posterior
distributions (solid lines) on the base parameters as listed in Section II.
Also displayed is the mean likelihood of the binned posterior samples (dotted 
lines).}
\end{figure}

\section{Observational constraints}
\label{sec:obsconst}

Turning first to the constraints on the basic parameter set, from 
Figure~\ref{1D_constraints} we note the overall consistency of our
results with the current observational picture (see for example Ref.~\cite{wmap}
and a work by two of the current authors Ref.~\cite{ll}), finding 
the $99\%$ marginalized probability regions to be 
\begin{eqnarray}
0.021<\Omega_{{\rm B}}h^2<0.027, & &0.10<\Omega_{{\rm C}}h^2<0.15\,,\nonumber\\
61<H_0<80,& &0.57<{\mathcal Z}<0.97 \,, \\
0.92<n_{{\rm S}}<1.07,& &19<A_{{\rm S}}<33 \,.\nonumber
\end{eqnarray}
%0.021<\Omega_{{\rm B}}h^2<0.026, & &0.1<\Omega_{{\rm C}}h^2<0.14\,,\nonumber 
%\\
%60<H_0<77,& &0.6<{\mathcal Z}<0.95 \,, \\
%0.92<n_{{\rm S}}<1.05,& &17<A_{{\rm S}}<30 \,.\nonumber
Note that part of our allowed region lies outside the priors assumed
by Nagata et al.~\cite{NCS}. As usual for joint analyses of CMB and
galaxy power spectrum data, it is
unnecessary to impose a further constraint on $H_0$.

\begin{figure}[t]
\includegraphics[width=\linewidth]{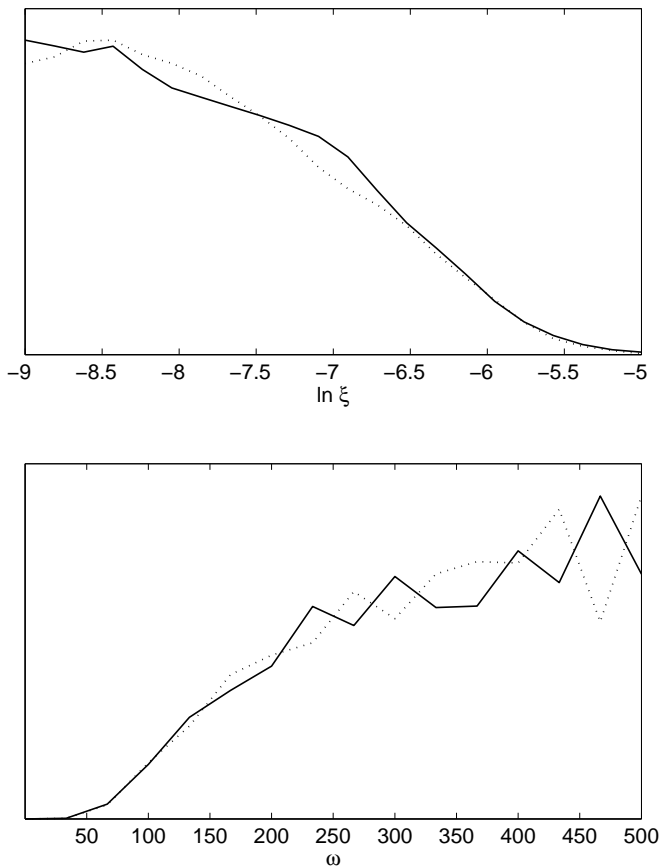}
\caption[1D_xi]{\label{1D_xi_omega} Marginalized 1D posterior
distributions (solid lines) on the BD parameter $\ln \xi$ (upper 
panel).
Also displayed are the derived importance sampled constraints (correcting for 
the
change in prior density) on the more familiar
$\omega$ (lower panel, no smoothing). We obtain a $95\%$ marginalized
probability bound of $\ln \xi > -6.2$, corresponding to a bound on the
BD parameter $\omega >120$.}
\end{figure}

The primary focus of our study has been to derive constraints on the
BD parameter for which, from the outset, we have expected
only to find a one-sided bound; the situation can only become more interesting
when both the angular diameter distance and the recombination history become
much better probed by the CMB. This expectation is indeed confirmed by the
data, as shown in Figure~\ref{1D_xi_omega} in which we display the region
of highest posterior density. The lower panel detailing the
posterior constraint on $\omega$ has been obtained by importance
sampling to correct for the change in prior density when changing
parameters from $\ln \xi$ to $\omega$ (we note that the mean
likelihood of the binned posterior obtained from sampling $\ln \xi$
performs well for putting a bound on $\omega$, demonstrating less
sensitivity to the details of the prior density).

 We obtain calculate marginalized probability upper bound and the
   main  result of this paper to be
\begin{eqnarray}
\ln \xi &<&-6.2,\;\;\; 95\%,\nonumber\\
\ln \xi &<&-5.7,\;\;\; 99\%.
\end{eqnarray}
The corresponding marginalized probability lower bounds on the  BD
  parameter are found to be 
\begin{eqnarray}
\omega &>&120,\;\;\; 95\% ,\nonumber\\
\omega &>&80,\;\;\;\;\; 99\%.
\end{eqnarray}
This bound is nicely consistent with the expectation for WMAP given by the 
Fisher
matrix analysis of Ref.~\cite{ck}.

\begin{figure}[t]
\includegraphics[width=\linewidth]{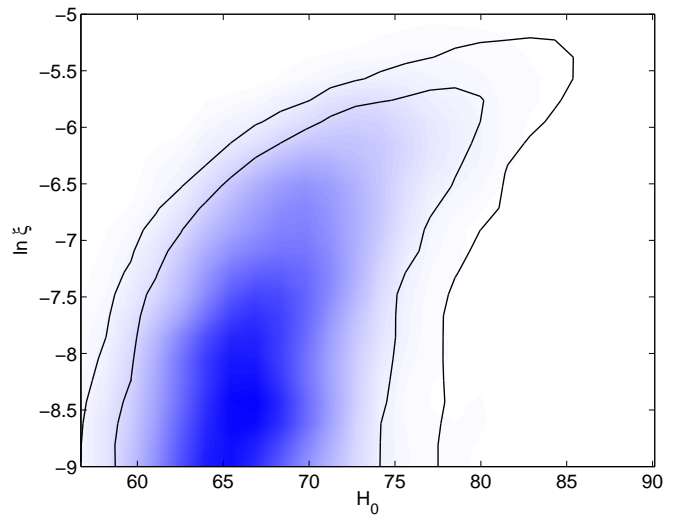} \\
\caption[H0xi]{\label{H0xi}Marginalized 2D posterior distribution in the
$\ln \xi$--$H_0$ plane. The solid lines enclose 95$\%$ and 99$\%$
of the probability. Under this parametrization there is clearly
a geometrical degeneracy.}
\end{figure}

\begin{figure}[t] \includegraphics[width=7.5cm]{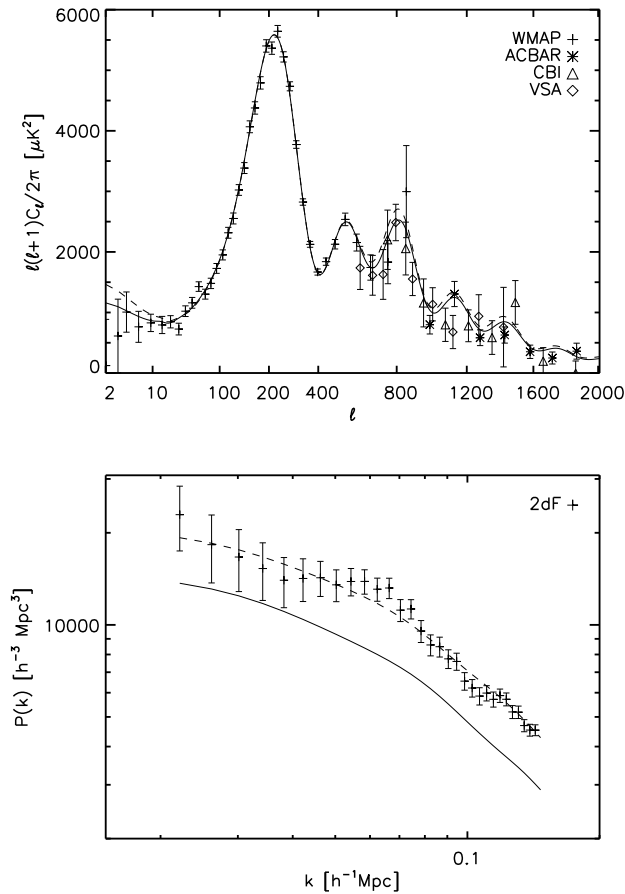} \\
\caption[cmb_matter]{\label{cmb_matter}A comparison between a $\Lambda$CDM model
(solid line) and a JBD $\Lambda$CDM model with $\omega =70$ (dashed line).  The
data are the 2dF galaxy power spectrum and the models the matter power spectrum
convolved with the 2dF window functions, and whose overall amplitude is left as
a free parameter.  Detailed parameters are given in Section~\ref{sec:obsconst}.}
\end{figure}

We present in Figure~\ref{H0xi} the 2D posterior constraints in the
$\ln \xi$--$H_0$ plane, in order to demonstrate the degeneracy and
covariance between these two parameters. In a more refined analysis, one
could replace $H_0$ with the dimensionless parameter $r_{{\rm s}}/D_{\rm A}$ 
more appropriate to the study of the CMB, where $r_{{\rm s}}$ is the sound 
horizon
at recombination and $D_{{\rm A}}$ is the angular diameter distance to the
last-scattering surface \cite{kosowsky}. Finally, in
Figure~\ref{cmb_matter} we display two models,
our best-fit  $\Lambda$CDM
model with parameters $\theta$$\equiv$$\{\Omega_{{\rm B}}h^2,
\Omega_{{\rm C}}h^2,H_0,{\mathcal Z}, n_{{\rm S}}, 10^{10}A_{{\rm S}},
\omega\}$$=\{0.023,0.12,66,$ $0.79,0.96,23.2,\infty \}$,
and a best-fit JBD model with parameters 
$\theta$$=\{0.024,0.13,79,0.80,1.03,24,70\}$, 
in order to illustrate how the observables change at finite
$\omega$. Here the JBD model lies in the
vicinity the contour enclosing $99\%$ of the posterior probability
distribution and was selected by running a short Monte Carlo
exploration at fixed $\omega = 70$.
Note that although in principle the parameter $\ln \xi$ could be
extended
to $-\infty$, whereby the bulk of the parameter space would be
composed of the $\Lambda{\rm CDM}$ model, in practice it is
reasonable to adjust the lower cutoff to the point where the
likelihood function loses sensitivity to 
the variation of $\ln \xi$ so that the Brans--Dicke model
alone is explored by the MCMC. Consequently, the probability
contours can reasonably be interpreted to describe the most
credible region of the  Brans--Dicke model parameter space.

 Our current analysis leaves the bias parameter free, and so
constrains only the shape of the matter power spectrum.  We note however that
the JBD model has a significantly higher amplitude, indeed requiring a modest
antibias $b \simeq 0.98$, which at least in part is due to the more rapid
perturbation growth ($\delta \propto a^{1+1/\omega}$ during matter domination
\cite{LMB}) in the JBD theory.  For comparison the $\Lambda$CDM model has a
best-fit bias $b = 1.2$.  This suggests that precision measures of the
present-day matter spectrum amplitude, as for instance may become available via
gravitational lensing, could significantly tighten constraints.  We also note
that there is a shift in the location of the baryon oscillations in the matter
power spectrum as compared to the $\Lambda$CDM model; these are mostly erased by
the 2dF window function,\footnote{Our analysis used 2dF data from Percival et
al.~\cite{2df}, preceding the more recent 2dF data analysis which shows evidence
of baryon oscillations \cite{Cole}.  We would not expect inclusion of this new
data (not yet publicly available) to significantly change our results.}  but
future high-precision measurements of those may also assist in constraining
$\omega$.

We have carried out the same analysis including also the data
from VSA, CBI and ACBAR in the multipole range $600<\ell <2000$. This 
high-$\ell$
data  leads to a slightly tighter bound on the
%Brans--Dicke parameter, $\ln \xi < -6.0$ corresponding to $\omega>100$.
Brans--Dicke parameter, $\ln \xi < -6.4$ corresponding to $\omega>177$
at $95\%$ marginalized probability. However, at the same time
inclusion of this new data leads to an unexpectedly large shift in the
spectral index, to $0.90<n_{{\rm S}}<1.00$ at $95\%$ marginalized
probability, so that the Harrison--Zel'dovich spectrum is only just
included (this statement remains true in the general relativity
limit).  Whether this points to some emerging tension in the combined
dataset, a harmless statistical fluctuation, or a hint of the breaking
of scale-invariance, can be addressed only in the light of the next
round of CMB observations. While our constraint on $\ln \xi$
marginalizes over $n_{{\rm S}}$, in the interests of quoting a robust
bound we have given as our main result the weaker limit obtained
without including the high-$\ell$ data.

Our ultimate constraint $\omega >120$ can be compared with that of Nagata et 
al.~\cite{NCS}, who quote results corresponding to $\omega>1000$ at two-sigma 
and $\omega>50$ at four-sigma. The former constraint is much stronger than 
projected in Ref.~\cite{ck}, and stronger 
than one would expect 
from a na\"{\i}ve assessment that the corrections to observables should be of 
order $1/\omega$. If we plotted a model with $\omega=1000$ in our 
Figure~\ref{cmb_matter}, it would lie practically on top of the $\Lambda$CDM 
model. However their latter constraint is in reasonable agreement with ours, and 
they do highlight that it is this constraint which corresponds to a sharp ridge 
of deteriorating chi-squared in their analysis, indicating that their constraint 
should conservatively be taken as $\omega>50$.

\section{Conclusions}

We have derived a constraint on Jordan--Brans--Dicke gravity from current 
cosmological observations, including cosmic microwave background (CMB) 
anisotropy data and the galaxy power spectrum data. Our main result is
to obtain a $95\%$ marginalized probability lower bound on the Brans--Dicke 
parameter
\mbox{$\omega >120$}. This result is complementary to the very strong
Solar System limit provided by Cassini, $\omega >40000$, as it probes
entirely different length and timescales. Our analysis is based on a Markov 
Chain Monte Carlo technique varying the basic cosmological parameters and 
$\omega$. 

At the present precision level, the greatest part of the constraining power 
comes from the shape of the CMB acoustic peaks, in particular from the 
first-year 
observations of WMAP. 
Therefore, assuming an extension to four years of the WMAP observations, 
we expect some further improvement on the limit on 
$\omega$ from cosmology. Further help is also expected from other 
structure formation data, as they improve quality and precision in 
coming years. In particular we have highlighted that an accurate measure of the 
present-day matter power spectrum amplitude, for instance from gravitational 
lensing, may be powerfully constraining when compared to the primordial 
amplitude from the CMB.

A leap forward in this and other contexts is expected from the observations of 
the 
Planck Surveyor probe, to be launched in 2007. Those observations are expected 
to be 
cosmic variance limited for the whole spectrum of CMB temperature anisotropy 
down to the damping tail, and to provide an accurate measurement of the gradient 
mode of the 
CMB polarization and its correlation with total intensity up
to the sixth acoustic peak \cite{DH}. According to the forecasts of Chen and 
Kamionkowski \cite{ck}, the limit on $\omega$ from Planck should be around an 
order of magnitude stronger than that from WMAP, and hence vastly stronger than 
the nucleosynthesis constraint. Whether that improvement can be realised from 
actual Planck data, of course, remains to be seen.

%======================================%
%<<<<<<<<<<< ACKNOWLEDGMENTS >>>>>>>>>>%
%======================================%

\begin{acknowledgments}

V.A.~was supported at Sussex by a Marie Curie Fellowship of the European
Community programme HUMAN POTENTIAL under contract HPMT-CT-2000-00096,
C.B. and F.P.~in part by NASA LTSA grant NNG04GC90G,
S.M.L.~in part by the European Union CMBNET network at Geneva, and A.R.L.~by
PPARC.  S.M.L.~thanks the University of Geneva for hosting the computations in
this work, and C.B.~and S.M.L.~acknowledge visits to Sussex supported by
PPARC.  We thank John Barrow for useful discussions, and acknowledge the use of
the Legacy Archive for Microwave Background Data Analysis (LAMBDA).  Support for
LAMBDA is provided by the NASA Office of Space Science.  
\end{acknowledgments}

%======================================%
%<<<<<<<<<<<< BIBLIOGRAPHY >>>>>>>>>>>>%
%======================================%

%%%%%%%%%%%%%%%%%%%%%%%%%%%%%%%%%%%%%%%%%%%%%%%%%%%%%%%%%%%%%%%%%%%%%%%%
\end{document}